\documentclass[11pt, oneside]{article}
\usepackage{geometry}
\usepackage[parfill]{parskip}
\usepackage{graphicx}	
\usepackage{amssymb}
\usepackage[]{authblk}

\usepackage{hyperref}
\usepackage[inline]{enumitem}
\title{\vfill Verbal Autopsy in Civil Registration and Vital Statistics: \protect \\ \textit{The Symptom-Cause Information Archive}}
\date{\today \vfill}

\author[1,2,*]{Samuel J. Clark}
\author[3]{Martin W. Bratschi}
\author[3]{Philip Setel}
\author[4]{Carla Abouzahr}
\author[5]{Don de Savigny}
\author[6]{Zehang Li}
\author[7,8]{Tyler McCormick}
\author[2,9]{Peter Byass}
\author[10]{Daniel Chandramohan}

\affil[1]{Department of Sociology, The Ohio State University, Columbus, OH, USA}
\affil[2]{MRC-Wits Rural Public Health and Health Transitions Research Unit (Agincourt), \protect\\School of Public Health, Faculty of Health Sciences, \protect\\University of the Witwatersrand, Johannesburg, South Africa}
\affil[3]{Vital Strategies, New York, NY, USA}
\affil[4]{Independent Consultant, Geneva, Switzerland}
\affil[5]{Swiss Tropical and Public Health Institute, University of Basel, Basel, Switzerland}
\affil[6]{Department of Biostatistics, Yale School of Public Health, New Haven, CT, USA}
\affil[7]{Department of Statistics, University of Washington, Seattle, WA, USA}
\affil[8]{Department of Sociology, University of Washington, Seattle, WA, USA}
\affil[9]{Department of Epidemiology and Global Health, Ume{\aa} University, Sweden}
\affil[10]{London School of Hygiene and Tropical Medicine, London, UK}

\affil[*]{Correspondance to \texttt{\href{mailto:work@samclark.net}{work@samclark.net}}, +1 (206) 303-9620}

\begin{document}
\pagenumbering{roman}
\maketitle
\clearpage
\pagenumbering{arabic}

\section{Verbal Autopsy in Civil Registration and Vital Statistics}

Poorly functioning civil registration and vital statistics (CRVS) systems are associated with poor population health due to inadequate planning and inefficient monitoring of progress. \cite{phillips2015well}  More than half of global deaths are not registered, \cite{mikkelsen2015global, wang2017global, adair2018estimating} 
and since then there has been renewed interest and investment in strengthening CRVS systems. \cite{bloomberg2015understanding,bloombergD4H,worldbankWhoCRVS}  These initiatives aim to increase the number and completeness of births and deaths that are registered and record a cause of death for as many deaths as possible.  Most of the unregistered deaths occur at home, leaving little trace in either clinical records or the civil register.  Assigning a cause of death to these `community' deaths is difficult, and to date the only feasible approach is verbal autopsy (VA). \cite{de2017integrating,garenne2014prospects,soleman2006verbal,setel2005sample}  

VA uses information obtained from the family and caregiver(s) of a decedent to assign a likely cause(s) of death.  Following the occurrence of a death and the identification of a suitable respondent, conducting a VA involves three components: 
\begin{enumerate*}[label=(\roman*)]
\item a questionnaire,
\item an interview, and
\item analysis of resulting data to assign a cause(s) of death.
\end{enumerate*}
For VA to be useful in routine mortality surveillance and CRVS, each of these must be standardized so that the resulting causes are derived through a consistent and replicable process that yields comparable causes.  This will ensure that cause of death distributions can be compared across time and among regions, thereby enabling a range of powerful epidemiological investigations that can inform public health planning and evaluation.  

The WHO Reference Group on VA has developed several standard VA instruments, \cite{whoStandardTools2012,whoStandardTools2016} and the most recent is compatible with all widely-used VA cause of death coding algorithms. \cite{nichols20182016}  Although VA interviews are conducted in a wide variety of sociolinguistic settings, comparatively little has been done to understand the effects of interview design or how to standardize the VA interview so that interviews conducted in different languages and cultural settings produce comparable data.  In contrast, there has been intensive work to develop and refine VA algorithms.

VA interviews produce mixed binary, quantitative, and narrative text data.  Standard practice has been to have physicians read VA interviews and assign causes of death. \cite{fottrell2010verbal} Physician-coding of VA (PCVA) is subject to physician-specific bias, occupies valuable physician time, is expensive, and  often leads to long delays between the VA interview and assigning a cause of death.  Over the past decade computational algorithms \cite{mccormick2016probabilistic,miasnikof2015naive,serina2015improving,serina2015shortened,byass2012strengthening,james2011performance,king2008verbal,king2010designing} have been developed and are increasingly used to interpret VA data and assign causes to VA deaths.  Algorithms are consistent, essentially cost-free (compared to the traditional use of physician time to assign cause of death to VA), and can be run on large numbers of deaths quickly.  Consequently, in addition to physician coding, VA algorithms are a potentially attractive option for routine mortality surveillance and CRVS applications.  However given the same set of signs and symptoms, different algorithms often assign different causes, and the literature does not provide clear guidance on which algorithm performs best or even how best to compare the performance of algorithms. \cite{desai2014performance,murray2014using,mccormick2016probabilistic} 

\section{Automated Cause Assignment Algorithms for Verbal Autopsy}

VA algorithms rely on three components:
\begin{enumerate*}[label=(\roman*)]
\item {VA data collected using a questionnaire},
\item {symptom-cause information (SCI)}, and
\item {a logical algorithm that combines the two to identify cause-specific mortality fractions (CSMF)s and/or assign a likely cause to each death}.
\end{enumerate*}
The SCI describes how VA symptoms are related to each cause.  In some cases SCI take the form of deaths with symptoms and causes assigned through another mechanism, and in other cases, the relationships are solicited directly from experts. There are six VA-coding algorithms that have been proposed and/or used widely: 
\begin{enumerate*}[label=(\roman*)]
\item \textit{InterVA}, \cite{byass2019integrated, byass2012strengthening, Fottrell2007,ISI:000208221400021,ISI:000235757400013,fottrell2011probabilistic}
\item \textit{Tariff}, \cite{james2011performance}
\item a derivative of Tariff called \textit{SmartVA-Analyze}, \cite{serina2015improving,serina2015shortened}
\item \textit{InSilicoVA}, \cite{mccormick2016probabilistic,mccormick2016probabilisticSupplement}
\item a naive Bayes classifier called \textit{NBC}, \cite{miasnikof2015naive} and 
\item the \textit{King-Lu} algorithm. \cite{king2008verbal,king2010designing} 
\end{enumerate*}  The list of causes that each algorithm assigns varies slightly.  InterVA and InSilicoVA can both assign causes defined by the WHO 2012 and 2016 VA standards, and importantly, these include maternal causes.  Tariff assigns causes defined by the Population Health Metrics Research Consortium (PHMRC), \cite{murray2011population} and SmartVA-Analyze assigns a subset of those causes (excluding maternal causes) using a shortened version of the PHMRC questionnaire -- the `PHRMC Shortened Questionnaire '\cite{serina2015shortened}.  King-Lu uses its own cause list and identifies CSMFs but does not assign causes to individual deaths. Significant work is underway to ensure that all algorithms can run using data from both the WHO 2016 VA standard questionnaire and the PHMRC Shortened Questionnaire. \cite{nichols20182016}  Open source, freely-available software\cite{openVA2019,smartVA2019,interva5} exists to run all of the algorithms.


Three recently published papers conduct comprehensive comparisons of the performance of VA algorithms and present contradictory results.  Desai et al. \cite{desai2014performance} compare algorithm-assigned causes to physician-assigned causes with all algorithms using their default SCI.  The algorithms performed poorly at replicating physician-assigned causes, and Tariff performed slightly better than InterVA4.  Murray et al. \cite{murray2014using} compare algorithm and physician-assigned causes.  In that comparison all algorithms used their default SCI and comparisons were done using the PHMRC `gold standard' dataset \cite{murray2011population} that contains deaths with VA and medically certified causes of death from six sites (all hospitals) in four countries during the early 2000s.  Their results show Tariff performing better than physicians, and InterVA4 performing worse than physicians.  Finally, the InSilicoVA Team \cite{mccormick2016probabilistic} compare algorithm-assigned causes. Uniquely, this comparison controlled for the effects of SCI in the comparisons; all algorithms used SCI derived from the PHMRC `gold standard' dataset, and all algorithms were trained using data from either all or single sites within the PHMRC dataset.  Controlling for SCI, InSilicoVA generally performed better than all the other algorithms, and InterVA4 performed better than Tariff. Moreover using a cross validation, multisite, train-test setup within the PHMRC dataset, the InSilicoVA Team clearly demonstrate that \textit{all} algorithms performed much less well when trained and tested on deaths from different sites within the PHMRC `gold standard' dataset (see the SCI-controlled comparison presented in the Supplementary Materials \cite{mccormick2016probabilisticSupplement}).  \textit{This strongly suggests that the SCI, and the information used to derive them, are more important than the logic of all existing algorithms}.  

Although there is no consensus among these results, collectively they are still useful and suggest a way forward.  With regard to underlying logic, the consensus is that some form of Bayes rule is the most fruitful approach, and with the sole exception of the Tariff family of methods, all algorithms use Bayes rule.  Except for the InSilicoVA team's work, all published comparisons conflate the effects of SCI and algorithm logic, and that makes their results difficult to interpret.  Using an ANOVA to apportion the variance explained by SCI and algorithm logic in a set of experiments run using the PHMRC `gold standard' dataset, the InSilicoVA Team demonstrates that the effects of SCI generally far outweigh differences in algorithm logic.\cite{clark2018quantifying}  Following directly from this, it is clear that \textit{a concerted effort must be made to significantly and rapidly improve the SCI available for VA algorithm developers and users.}  Finally, a standard design for comparison and a standard set of performance metrics would greatly improve the comparability and interpretability of VA algorithm performance comparisons.

\section{Recommendation: the Symptom-Cause Information Archive}

Because SCI relates VA symptoms to causes of death, the easiest and most reliable way to obtain SCI is from deaths with a VA and a cause(s) assigned through an independent mechanism. These `labeled' deaths can be processed to produce a quantitative description of how each symptom or group of symptoms is related to each cause.  In a Bayes rule framework this typically consists of conditional probabilities of observing a symptom or group of symptoms when death results from a specific cause, $\Pr(s|c)$ or $\Pr({s_1, \dots, s_j}|c)$.  The independent mechanism for assigning causes to the `labeled' deaths can be 
\begin{enumerate*}[label=(\roman*)]
\item medical certification of cause of death, or
\item a standard autopsy, or
\item physician-assigned causes using the VA data, or 
\item a variety of mixtures of the two with perhaps additional information from clinical records and/or minimally-invasive tissue samples. \cite{bassat2017validity,castillo2015pathological,castillo2016validity}
\end{enumerate*}

To allow VA algorithms to be used with confidence in routine mortality surveillance and CRVS systems in multiple settings and as time progresses, it is necessary to have SCI based on deaths from a wide variety of settings that accumulate continuously.  This variety can be achieved by continuously pooling deaths with VA and an independently-assigned cause(s) (e.g. by performing PCVA for a portion of VAs) from various settings and updating SCI based on those deaths on a regular schedule.  Then by definition, the resulting SCI will be both widely representative and continuously updated.  This will ensure that the SCI used to code a specific death represents both the broad relationships between symptoms and causes across a diversity of settings and the specific relationship between particular symptoms and causes at the time and place of that specific death.  The InSilicoVA Team's results\cite{mccormick2016probabilisticSupplement,clark2018quantifying} strongly suggest that this will greatly improve the performance of \textit{all} VA algorithms to identify both CSMFs and assign causes to individual deaths. 

The practical implementation of such an idea is a centralized archive of deaths with VA and an independently-assigned cause(s) that continuously accepts deaths from a widely dispersed group of collaborators and periodically updates and disseminates SCI based on those deaths -- \textit{the SCI Archive}.   

A common desire expressed by VA users in routine mortality surveillance and CRVS systems is to have VA algorithms `replicate what \textit{our} physicians would do' with VA data.  This replication can be achieved by using SCI based on deaths with physician-coded VAs representative of the area where the VAs are conducted.  Physicians code VA deaths with their own idiosyncratic bias which is one of the primary motivations for developing algorithms.  This bias can be quantified and eliminated in the preparation of SCI from PCVAs provided two conditions are met: \cite{mccormick2016probabilistic}
\begin{enumerate*}[label=(\roman*)]
\item each VA death is read and coded by at least two physicians and 
\item each of the coding physicians reads and codes many VAs.
\end{enumerate*}
Physician-coded VAs that meet these conditions are `de-biased PCVAs' (DBPCVA).  

A way to quickly establish the SCI Archive would be to start with DBPCVA deaths contributed by any VA practitioner willing to provide them.  The resulting SCI would enable VA algorithms to assign causes in a way that replicates what non-biased physicians would do.  

\bibliographystyle{unsrt}
\bibliography{sci-va}

\end{document}